# HugSense: Exploring the Sensing Capabilities of Inflatables

Klaus Stephan
Chemnitz University of Technology
Germany
klaus.stephan@informatik.tu-chemnitz.de

Maximilian Eibl
Chemnitz University of Technology
Germany
eibl@informatik.tu-chemnitz.de

Albrecht Kurze
Chemnitz University of Technology
Germany
albrecht.kurze@informatik.tu-chemnitz.de

## Abstract
What information can we get using inflatables as sensors? While using inflatables as actuators for various interactions has been widely adopted in the HCI community, using the sensing capabilities of inflatables is much less common. Almost all inflatable setups include air pressure sensors as part of the automation when pressurizing or deflating, but the full potential of those sensors is rarely explored. This paper shows how to turn a complete pillow into a force sensor using an inflatable and a simple pneumatics setup including an air pressure sensor. We will show that this setup yields accurate and interesting data that warrants further exploration and elaborate on the potential for practical applications.

## 1 Introduction

Recently, easily replicable methods for rapid prototyping of inflatables like WireShape from Gohlke et al. [3] have become available, making the creation of inflatables for tangible interfaces, data physicalization, shape-changing interactions, and other applications more accessible. Most of the design artifacts created are actuated inflatables like the inflatable flowers in Danilisana et al. [1], where form change is the main focus. Other applications like the Force Jacket developed by Delazio et al. [2] focus on the haptics of actuated inflatables. There currently is no real focus on exploring the sensor data that an inflatable can provide.

Most pneumatic setups have integrated air pressure sensors for the automation of pressurization and deflation. Even though these sensors are only used to measure when certain pressure thresholds are reached, they have a very high precision that goes far beyond what would be necessary for this application. This makes it possible to use the same pneumatic setups to turn inflatables into force sensors if they are not in the process of pressurization or deflation.

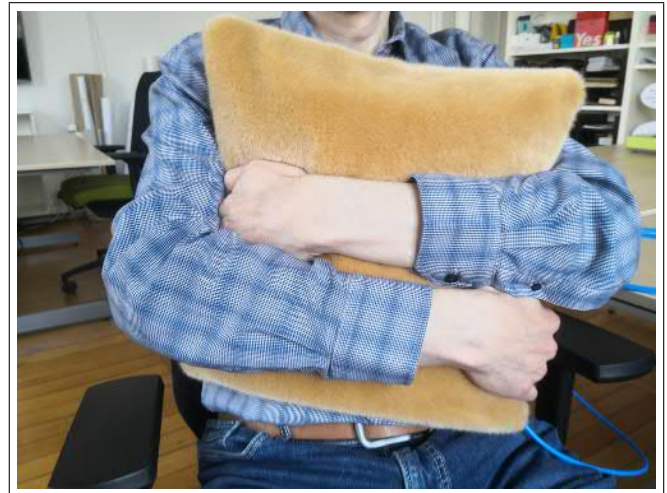

**Figure 1: Usage of the HugSense**

Our approach uses inflatables and a measurement of their internal pressure to turn objects into force sensors, which is much easier and more affordable than covering large objects like a pillow in flexible force sensitive resistors or other force sensors. Even though this reduces the interaction to a single sensor value, the high precision of the air pressure sensor makes it possible to discern a variety of interactions and effects like respiration or jerks of the user while hugging the pillow, general activity when playing with the pillow or specific patterns like the activity of the air pump.

## 2 Build

The HugSense (Figure 2) consists of two units: a control unit and the pillow.

*Control Unit.* The control unit has two buttons for inflating and deflating the pillow. It contains an ESP32-S3 microcontroller board and a pneumatic setup that includes a Honeywell MPRLS0025PA00001A air pressure sensor. The inside of the control unit can be seen in Figure 4. The the paper "Pressure Proof: A DIY Guide to Creating a Pneumatic Testing System for Validating Inflatable Product Designs" from Stephan et al. [7] provides a detailed build guide of the pneumatic setup.

*Inflatable Pillow.* The inflatable pillow consists of a plush pillowcase and an inflatable pillow made from TPU coated Nylon with the WireShape method outlined in Gohlke et al. [3]. Both can be seen in Figure 3. The pillow is ca. 40 cm wide and square, giving

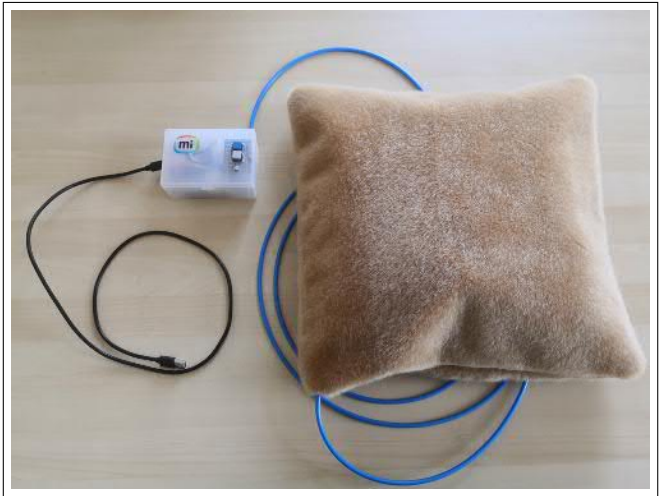

**Figure 2: The HugSense**

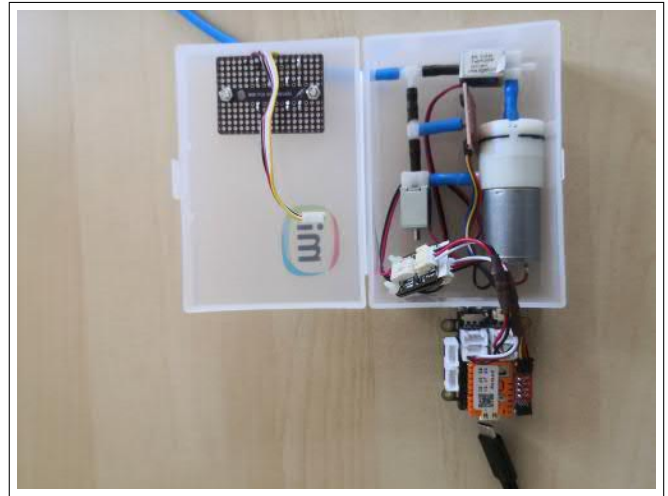

**Figure 4: Control Unit of the HugSense**

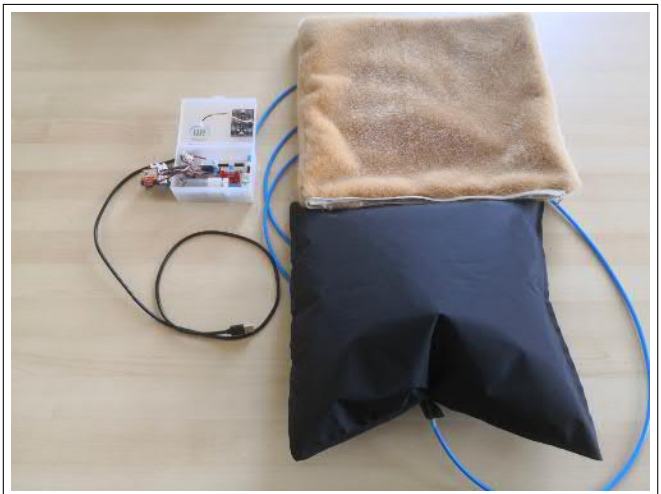

**Figure 3: Overview of the Parts of the HugSense**

it an appropriate size to be comfortably hugged by an adult.

A notable difference between the usage of inflatables as sensors compared to use as actuators is that the volume of the inflatable is much less of a concern. While in an actuator, the throughput of the air pump limits the size of an inflatable that can be pressurized fast enough for the intended interaction, an inflatable used as a sensor can be prefilled without time constraints. As pressure changes travel with the speed of sound, the impact of size on the reaction time of the sensor is negligible even when larger inflatables are used.

To keep the pillow of the HugSense soft in our preliminary tests, we prefilled it just enough to prevent opposite sides of the pillow from touching when the pillow is squeezed, which is around 75% of the total air capacity. This also means that the pillow is not pressurized significantly above atmospheric pressure unless it is squeezed by the user, giving us an idle state with just above 95500 Pa pressure.

## 3 Usage and Capabilities

As can be seen in Figure 1, the HugSense pillow is intended to be held in a hug. If it is held in this position, it is possible to interpret the data of the pressure sensor since there is enough context for what is happening to the pillow. If the pillow is freely handled, the interpretation of the data is much harder, but might be possible with additional information or by creating a set of actions that should be performed on the pillow that could then be recognized.

The plot in Figure 5 shows the output of one of our initial tests. The annotated sections represent the following:

(1) The pressure increase from not interacting with the pillow to hugging it lightly.
(2) Remaining still while hugging the pillow, we can see the respiration of the user in small but periodic pressure changes.
(3) Pressure spikes occur during jerking motions when the user is surprised, startled or scared.
(4) Shows the output of random interactions with the pillow, e. g. when cuddling with it without hugging.

Those tests show, that there is potential to generate useful data with the HugSense pillow. As the pillow is easy to use and doesn't require a lot of setup from the users side, it could be a valuable inclusion for physiological measurement equipment in psychology, either adding data points that current equipment doesn't recognize or replacing other methods that require more setup or are more intrusive. This could be especially interesting when working with children or other vulnerable groups, where giving them a plushy pillow to hug is much less stressful than wiring them with electrodes and other sensors.

## 4 Future Work

Our ideas for future work are threefold:

First, it should be explored how the data from HugSense correlates to other physiological measurements and how it can be best utilized for psychological studies.

Second, we plan to explore how machine learning and other analytic approaches can be applied to the data and if it is possible to reliably identify other interactions and effects. This could broaden the application of inflatables as sensors significantly.

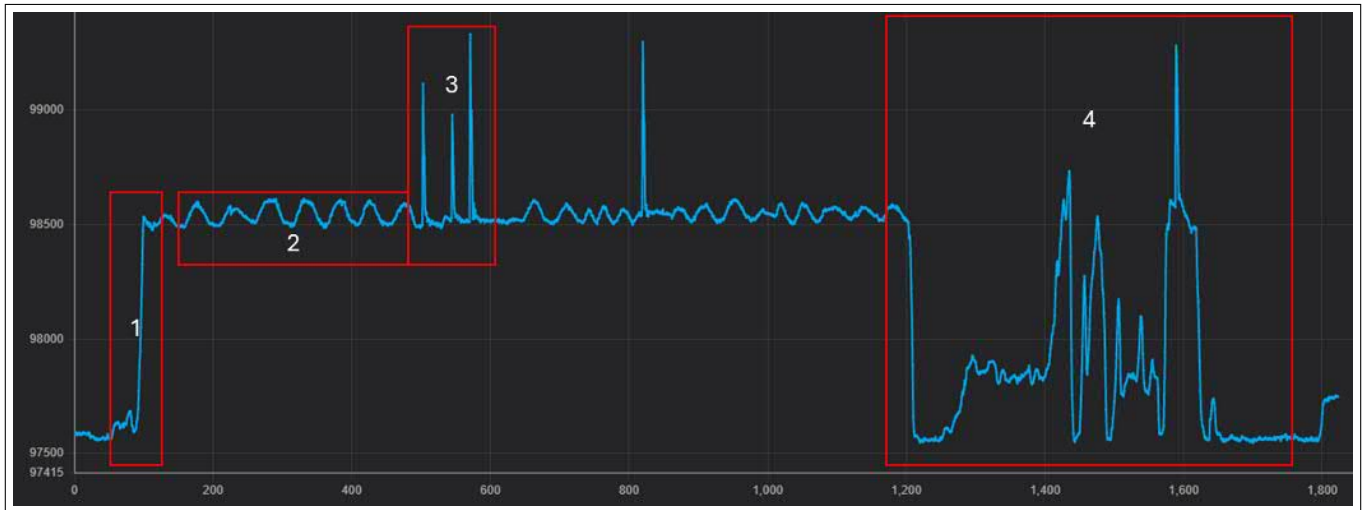

Figure 5: Plot of the Data HugSense provides

Lastly, it would be useful if the inflatable could simultaneously be used as actuator and sensor. This poses a challenge, because the pump and valves compromise the sensor data and because interactions that change the inflatables pressure make the control of the actuator unreliable. Discerning what changes in pressure originate from the actuating of the inflatable and what is caused by interactions is the key challenge for this use case.

In concrete use cases we imagine a sensor-actuator combination in our smart textiles co-design tool for workshops, the Wheel of Plush [5], to combine the already existing soft pneumatic cushion and a pressure sensor cushion.

Besides a use in the lab or in workshops we also see good chances for the use in actual products since the principle is simple and cheap to realize. Accordingly, we plan to transfer these insight together with our multi-dimensional normalized sensor-actuator mapping approach [6] also to smart soft toys [4] turning the whole soft toy into a sensor-actuator combination, intended for playful interaction and daily use.

## 5 Conclusion

The data from those initial tests shows, that even this simple setup yields interpretable data that might be useful in addition to established physiological measurements like pulse or skin conductivity. It might even be possible to partially replace other methods that require extensive wiring of the user in favor of just giving them a pillow to hug, which could be valuable when working with children or other vulnerable groups, where extensive preparations are too intrusive.

Overall, this shows that inflatables have interesting potential not only as actuators, but as sensors, and we think this should be explored further.

We invite the HCI community to try out and hope to inspire adaptation and new use cases.

## Acknowledgments

This research is funded by the German Ministry of Research, Technology and Space (BMFTR) grant FKZ 16SV9117.

## References

[1] Hanna Danilishyna, Alisa Popp, Rosa Koningsbruggen, and Eva Hornecker. 2024. Exploring Emotion physicalization Through Soft Robotics. In *Mensch und Computer 2024 - Workshopband*. Gesellschaft für Informatik e.V. doi:10.18420/muc2024-mci-demo-371
[2] Alexandra Delazio, Ken Nakagaki, Roberta L. Klatzky, Scott E. Hudson, Jill Fain Lehman, and Alanson P. Sample. 2018. Force Jacket: Pneumatically-Actuated Jacket for Embodied Haptic Experiences. In *Proceedings of the 2018 CHI Conference on Human Factors in Computing Systems* (Montreal QC, Canada) *(CHI '18)*. Association for Computing Machinery, New York, NY, USA, 1–12. doi:10.1145/3173574.3173894
[3] Kristian Gohlke, Hannes Waldschütz, and Eva Hornecker. 2023. WireShape - A Hybrid Prototyping Process for Fast & Reliable Manufacturing of Inflatable Interface Props with CNC-Fabricated Heat–Sealing Tools. In *Proceedings of the 8th ACM Symposium on Computational Fabrication* (New York City, NY, USA) *(SCF '23)*. Association for Computing Machinery, New York, NY, USA, Article 7, 8 pages. doi:10.1145/3623263.3623363
[4] Albrecht Kurze, Lewis Chuang, Arne Berger, Klaus Stephan, Natalie Sontopski, Stephan Hildebrandt, Elisabeth Jost, and Maximilian Eibl. 2023. Bitplush: Unleashing the Paws-ibilities of Smart Materials in Smart Plush Toys. Mensch und Computer 2023 - Workshopband. doi:10.18420/muc2023-mci-ws09-420
[5] Natalie Sontopski, Stephan Hildebrandt, Lena Marcella Nischwitz, Klaus Stephan, Albrecht Kurze, Lewis L Chuang, and Arne Berger. 2024. Wheel of Plush: A Co-Design Toolkit for Exploring the Design Space of Smart Soft Toy Materiality. In *Proceedings of the 2024 ACM Designing Interactive Systems Conference* (Copenhagen, Denmark) *(DIS '24)*. Association for Computing Machinery, New York, NY, USA, 2958–2971. doi:10.1145/3643834.3660744
[6] Klaus Stephan, Maximilian Eibl, and Albrecht Kurze. 2024. Just the Normal Fluff: The Wheel Of Plush as an Example for Human Centric Data Normalization. Mensch und Computer 2024 - Workshopband. doi:10.18420/muc2024-mci-ws02-160
[7] Klaus Stephan, Maximilian Eibl, and Albrecht Kurze. 2025. Pressure Proof: A DIY Guide to Creating a Pneumatic Testing System for Validating Inflatable Product Designs. Mensch und Computer 2025 - Workshopband. doi:10.18420/muc2025-mci-ws11-137